\documentclass[%
 aip,
 amsmath,amssymb,
 reprint,%
]{revtex4-1}

\usepackage{graphicx}% Include figure files
\usepackage{dcolumn}% Align table columns on decimal point
\usepackage{bm}% bold math
\usepackage[utf8]{inputenc}
\usepackage[T1]{fontenc}
\usepackage{mathptmx}
\usepackage{etoolbox}
\usepackage{amsmath} %required for \DeclareMathOperator
\DeclareMathOperator\erfc{erfc}

\makeatletter
\def\@email#1#2{%
 \endgroup
 \patchcmd{\titleblock@produce}
  {\frontmatter@RRAPformat}
  {\frontmatter@RRAPformat{\produce@RRAP{*#1\href{mailto:#2}{#2}}}\frontmatter@RRAPformat}
  {}{}
}%
\makeatother
\begin{document}

\preprint{AIP/123-QED}

\title[Preprint Draft: Direct Two-Dimensional Goniometric Steering of Vacuum Electrospray Ion Beams for Angular Time-of-Flight Studies]{Direct Two-Dimensional Goniometric Steering of Vacuum Electrospray Ion Beams for Angular Time-of-Flight Studies}
% Force line breaks with \\
\author{Zach Ulibarri}
 \email{zulibarri@cornell.edu}
\author{Elaine Petro}%

\affiliation{ 
Sibley School of Mechanical and Aerospace Engineering, Cornell University%\\This line break forced with \textbackslash\textbackslash
}%

\date{\today}% It is always \today, today,

\begin{abstract}
Here we present a novel method for two-dimensional steering of vacuum electrospray ionization beams to better understand their angular properties. Utilizing an externally wetted tungsten needle with the ionic liquid 1-Ethyl-3-methylimidazolium tetrafluoroborate (EMI-BF$_4$), we employ a dual-axis goniometer to achieve in-vacuo beam steering in both pitch and yaw. This setup enables detailed angular-dependent time-of-flight (TOF) measurements of molecular species within the ESI plume as a function of plume angle. Our findings reveal significant variations in the relative abundance of monomers, dimers, trimers, and heavy species across different beam angles, with monomers being at a relative minimum and fragments, trimers, and heavy species being at a relative maximum at the beam center. At higher angles, the relative monomer abundance increases, but at the plume extremities heavy species and trimer fragments begin to rise. The ability to accurately aim the ion beam dramatically enhances the signal-to-noise ratio for various diagnostic tools, underscoring the utility of this system for both scientific studies and practical laboratory applications. This two-dimensional steering capability offers a robust framework for future investigations into ESI plume dynamics and composition, enabling more precise characterizations that are critical for optimizing ESI-based propulsion systems and other applications.
\end{abstract}

\maketitle

\section{\label{intro}Introduction}
Vacuum electrospray ionization (ESI) sources are key components of electrospray thrusters, which are among the most precise and efficient spacecraft propulsion systems currently available \cite{krejci2018space,petro2020characterization,wirz2019electrospray,suzuki2021fabrication}. Also called colloidal thrusters, these electric propulsion (EP) sources use a high potential to accelerate ions from a room temperature ionic liquid (RTIL) at high velocity to produce thrust. Ionic liquids are ionically coupled salts, composed of an anion-cation pair. RTILs are ionic liquids that exist as liquids in ambient conditions. Many RTILs have negligible vapor pressures, enabling liquid operation in the vacuum of space.

Vacuum ESI systems are typically lifetime-limited due to degradation after long firing times, often thought to be the result of stray ions or droplets emitted at high-off-axis angles collecting on thruster or spacecraft surfaces and forming electrical shorts between high-voltage components \cite{thuppul2021mass,smith2024propagating,enomoto2022molecular}. For optimally efficient thrust, pure ion emission of the ionic liquid monomer (the anion in negative ion firing and the cation in positive ion firing) is desired, as this gives the maximum thrust yield per unit mass of fuel. However, in practice some fraction of the emitted current will come from combinations of the anions and cations \cite{lozano2005efficiency} called dimers and trimers, as well as higher mass species (e.g., droplets).

For the ionic liquid 1-Ethyl-3-methylimidazolium tetrafluoroborate (EMI-BF$_4$), the positive ion monomer would be EMI$^+$, the dimer would be (EMI-BF$_4$)EMI$^+$, and the trimer would be (EMI-BF$_4$)$_2$EMI$^+$. There would be similar complexes in the negative mode that substitute the anion BF$_4^-$ for the cation EMI$^+$. In addition, the dimers may fragment into monomers and a neutral, and the trimers may fragment into a dimer and a neutral, etc. When such fragmentation occurs within the acceleration column of a mass spectrometer, it will appear in mass spectra as having a mass-to-charge ratio (m/z) value in between the parent and daughter molecule species.

The molecular composition of vacuum ESI plumes are not well-understood, in particular with respect to these fragments produced in high-velocity impacts between plume constituents (or through unintentional contact with surfaces throughout the thruster), as these may then also move in off-axis directions to contribute to degradation or reduced performance \cite{shaik2024characterization,geiger2024secondary,miller2020measurement}. Understanding the angular-dependence of these interactions and the molecular species distribution is therefore critical for designing ESI propulsion sources with longer operational lifespans and increased efficiency \cite{smith2024propagating,bendimerad2024investigating,mier2017spacecraft}. Additionally, due to their unique properties, ESI systems are used in a wide variety of scientific \cite{chingin2009exploring,cogan2023electrospray,chiu2007vacuum}, industrial \cite{reneker1996nanometre,morato2023desorption}, and pharmaceutical \cite{loscertales2002micro} contexts. Thus, a greater understanding of the physics of vacuum electrospray ion emission and the downstream plume characteristics is desirable.

While vacuum ESI systems have been investigated in a variety of contexts, very few studies have considered the angular properties of vacuum electrospray plumes.  Experimental measurements are almost uniformly taken along the axis of the emitter tip, and in the absence of well-characterized angular plume properties, the beams are often taken to be somewhat homogeneous with respect to their molecular composition. \emph{Schroeder et. al 2023} \cite{schroeder2023angular} studied a vacuum ESI plume from a porous emitter tip on a rotational stage that allowed the beam angle to be varied in one dimension. A time-of-flight mass spectrometer based on a channel electron multiplier (CEM) detector was used to study the resultant plume. The results from the CEM were then collated to produce a simulated full-beam TOF curve, from which the relative compositions of the monomers, dimers, and trimers were measured as a function of accelerating voltage rather than angle.

Using a Kalman update algorithm \cite{jia2022quantification}, \emph{Jia-Richards 2024} \cite{jia2024numerical} analyzed the 800 V dataset from \emph{Schroeder et. al} to provide true angular-dependent compositional measurements. This study found that dimers dominated the center of the beam, monomers dominated the extremities, and that trimers and high-mass species were most prevalent at the center but weakly present everywhere. The compositional study yielded a center line composition of "approximately 34\% monomers, 48\% dimers, 9\% trimers, and 9\% heavy species," while the extremities exhibited "67\% monomers, 29\% dimers, 1\% trimers, and 3\% heavy species." \cite{jia2024numerical}

However, because \emph{Schroeder et. al} \cite{schroeder2023angular} was only capable of varying the plume angle along a single degree of freedom, there is no guarantee that the beam slice that was characterized was along the true center line of the emission beam. A key finding of original study was that the beam's center angle varied from +7$^{\circ}$  to -8$^{\circ}$ across an acceleration voltage range 800 to 950V, and the study found evidence of multiple emission sites at the highest potential \cite{schroeder2023angular}. With such a spread found along the one scanned dimension, there is no reason to believe that the two-dimensional conical beam was centered in the second, static dimension. That is, the beam slice that was measured in this study almost certainly would have been at an arbitrary angle in another dimension that does not pass through the true conical beam center. Indeed, the second emission site found at high potential in \emph{Schroeder et. al} was weakly identified and posited to be "off axis in the direction orthogonal to that investigated" \cite{schroeder2023angular}.

\emph{Petro et. al 2022} \cite{petro2022multiscale} created a multi-scale model to describe electrosprays across multiple modelling scale sizes, combining electrohydrodynamic fluid modeling with n-body particle dynamics to capture the evolution of an individual ion plume from the sub-micrometer scale emission region to the steady-state region further downstream. It also uses a fragmentation model informed by molecular dynamics simulations, finding increased collisions and fragmentation near the plume center where density is highest. This study also compared the theoretical results to experimental data\cite{lozano2005energy} where an ESI source was mounted on a rotating stage to enable 1-dimensional steering. The plume divergence angle from the experimental data was found to be slightly wider than the theoretical models, although the authors explain that this is likely due to discrepancies in the emitter geometries of the model and the experiment. Even so, the experimental data used in this study also necessarily could only be characterized in a single dimension, and so there is again no guarantee that the data was taken across the true conical beam center. Thus, there is a need for angular measurements taken across a beam slice known to be in the conical beam center, and there is a more broadly a need for the ability to study vacuum ESI plumes in two dimensions rather than just one.

Furthermore, because many emitters fire at least slightly off-axis, even when simply attempting to study the center axis of an ion plume, many experiments unknowingly sample off-center. While such off-center sampling complicates efforts to characterize and understand the plume dynamics and angular molecular variation, it may also dramatically reduce the signal-to-noise ratio (SNR) detected by downstream components, as they may sample lower-current regions of the emission plume. Thus, in addition to the need for understanding the angular characteristics of the plume, there is great laboratory utility in being able to simply and rapidly steer the beam onto a target or diagnostic tool to increase the system SNR. Such capability would dramatically improve a wide variety of scientific measurements in everyday laboratory experiments.

Here we present a simple and novel method of in-vacuo, two-dimensional direct aiming of a tungsten needle vacuum ESI source externally-wetted with the ionic liquid EMI-BF$_4$ to study the angular characteristics of the emission plume. Through the use of a dual-stage goniometer, rotational feedthroughs, flexible shaft couplings, and a simple downstream current-detecting electrode, ESI beams can be rapidly aimed to put the highest possible current on target to maximize system SNR for a variety of instruments. Importantly, the axis of rotation is at or very close to the emission site, so downstream ion optics or TOF gates can be placed in the beamline without concern that moving the emitter angle will change the desired beam center line. That is, pass-through instruments can be placed very close to the emitter, and even as the firing angle is moved across a few tens of degrees in two dimensions, the emission site will be held in a constant position so that the emitted ions will still pass through the center of those instruments.

By directly steering the emitter in both pitch and yaw, the firing angle was found to be wildly off-center, at 12.5$^{\circ}$ in pitch and 4.7$^{\circ}$ in yaw. The pitch was then held constant at 12.5$^{\circ}$ and the beam was scanned in yaw across the goniometer's maximum effective throw range of $\pm 23^{\circ}$ (with respect to the emitter needle's center axis). This method ensures that the beam is sampled in a one-dimensional slice along the true center line of the emission plume cone.

Positive ion TOF measurements were taken across the yaw scan, and the relative ion yields of the monomers, dimers, trimers, fragments thereof, and high mass species were directly measured from the TOF curves. We find that fragmenting trimers and high mass species are most prevalent near the center line, but that monomers and dimers dominate the signal everywhere. The goniometric system is useful for providing angular-dependent measurements of this sort, but it is also an exceptionally useful utility in the laboratory for all manner of ESI experiments.

\section{Experimental Setup}
\label{setup}

\begin{figure}
	\centering
		\includegraphics[width=1\linewidth]{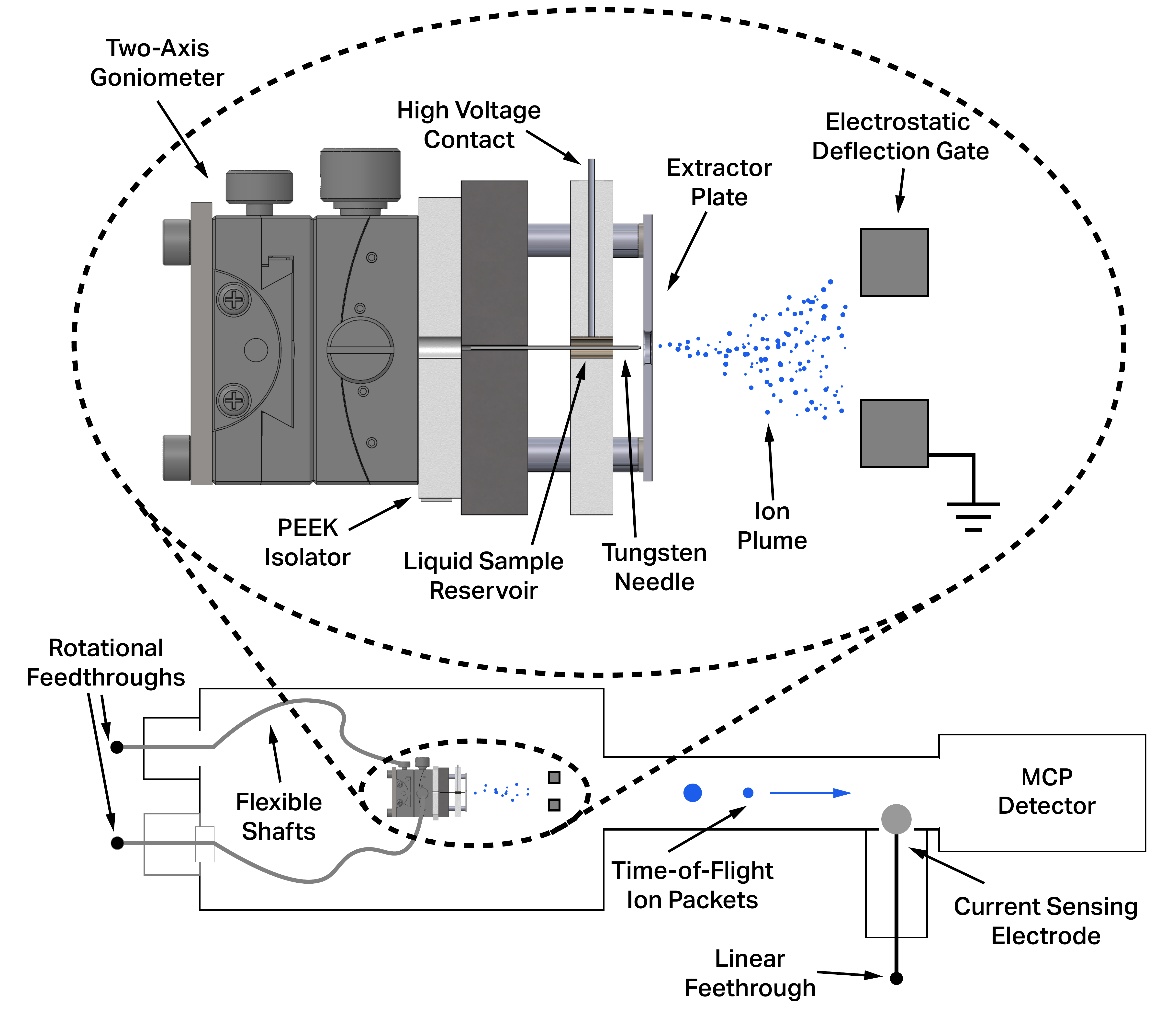}
	\caption{A diagram of the two-dimensional vacuum electrospray ionization (ESI) steering system. The ESI source is mounted on a dual-axis goniometer, and flexible shafts attached to rotational feedthroughs enable in-vacuo manipulation of the goniometer controls. The system may rotate in two dimensions about the emission site in the center of the extractor aperture. The ESI source is fired through an electrostatic gate towards a microchannel plate detector at the end of a flight tube to enable TOF operation. A current sensing electrode is mounted on a linear feedthrough that enables it to be rapidly inserted into or taken out of the ion beam line.}
	\label{sourcefig}
\end{figure}

\subsection{The vacuum ESI Source}

A diagram of the goniometric vacuum ESI system is shown in Fig. \ref{sourcefig}. A tungsten needle passes through a sample reservoir filled with the ionic liquid EMI-BF$_4$. The particular needle used in these experiments has an apex radius of about 3.5 $\mu$m after electrochemical etching. Scanning electron microscope images of the needle tip used in these experiments are shown in Supplemental Figs. \ref{needle1} and \ref{needle2}.

A solution of a few tens of $\mu$L is loaded into the stainless steel reservoir by use of a micropipette. The vacuum chamber is brought down to high vacuum, typically below 1 $\times 10^{-5}$ Torr with a turbomolecular pump. The acceleration potential, $\phi_{acc}$, is applied to the ionic liquid via the a metal bolt passing through the insulating teflon layer and contacting the steel reservoir to reduce electrochemical degradation of the needle \cite{brikner2012role}. A grounded stainless steel extractor plate with a circular aperture in the center is held just above the needle tip, with precision ground ceramic spacers used to ensure that it is held perpendicular to the needle tip. 

As the reservoir is brought up to $\phi_{acc}$, the potential between the ionic liquid and the extractor aperture creates a Taylor cone emission site and ions are accelerated downstream \cite{taylor1964disintegration,gamero2001electrospray,fernandez2007fluid}. A Keithley 2657A source meter unit (SMU) is used to supply $\phi_{acc}$ while providing high-precision output current measurement. High precision pico-ammeters are used to monitor the current on the extractor and on the gate grounding plates, described in Section \ref{TOF}. The extractor plate received small but negligible current when the ESI source was firing, and thus it was assumed that virtually all of the emitted current made it through the extractor. However, the amount of current intercepted by the gate plates was as high as 50$\%$ as the yaw angle was scanned. 

\subsection{Time-of-Flight}
\label{TOF}
The TOF system used in this experiment is further detailed in Ref. \citenum{cogan2023electrospray}, and so only a brief description is given here. Emitted ions travel through an electrostatic gate composed of two electrodes held parallel to each other. A fast circuit is used to drive these electrodes to 2 kV with respect to each other in a 1 kHz 50$\%$ duty cycle square wave with a rise time of 25 ns. Grounded grids with high transparency are used to encase the electrodes to prevent errant electric fields. When the gate potential is on, the ion beam is deflected into the chamber walls. When it is rapidly switched off, the ion beam is allowed to pass through the electrostatic gate towards the detector.

Many mass spectrometer systems use extremely fast pulses, such as laser pulse ablation (e.g, \cite{henderson2015direct,klenner2022developing}), to produce well-defined instantaneous ion emission. The rise time of the high voltage gate pulse in the system presented here precludes such operation, and the vacuum ESI sources produce startup transient currents when rapidly pulsed on or off. As such, this system is different from many MS systems in that it produces continuous current rather than fast current impulses. Thus, it does not yield sharp Gaussian mass spectral pulses, but a current yield that rises when new populations of ions begin arriving, as described in Ref. \citenum{cogan2023electrospray}. The derivative of this current yield can be used to create more traditional mass spectra to aid in analysis.

The ions that are allowed through the gate continue down a flight tube approximately 1 meter in length to reach a Hamamatsu F1217-011 microchannel plate detector (MCP) detector. The MCP's output current signal is fed to an ARICorp TDC-30 transimpedance amplifier with a gain of 0.5 V/$\mu$A. While the MCP's nominal amplification voltage is 1 kV, with the precise aiming afforded by 2-D goniometric system described below, the signal levels would often rise beyond the input range of the TDC-30 amplifier. Thus, for the experiments described here, the MCP was operated at a reduced amplification voltage of 800 V. The voltage signal from the TDC-30 is fed into an oscilloscope that is triggered by the square wave of the gate pulse, enabling thousands of spectra to be averaged to dramatically increase SNR. All the results presented here make use of this averaging, with one thousand or more triggers used in each recorded TOF curve. Due to the limitations of the setup, the averaging was done in the scope.

\subsection{The Goniometric Aiming System}

\iffalse
\begin{figure}
	\centering
		\includegraphics[width=1\linewidth]{images/gonivsrotation.png}
	\caption{A diagram showing the advantage of using a goniometer over a simple rotation stage. For high-off-axis emission plumes, simple rotation stages will cause the emission site to physically move as the beam is aimed. A goniometer which rotates about a critical point some distance above its surface. By machining the ESI source such that the emission site is at this critical rotation point, the beam can be directly aimed across a wide angular range in two dimensions while keeping the emission site in a constant physical location. This enables the emission beam to pass through the center of any downstream equipment, such as the electrostatic gates shown here. Importantly, the dual-axis goniometer enables such operation in both axes, enabling two-dimensional rotation about the emission site in-vacuo.}
	\label{gonivsrotation}
\end{figure}
\fi

The ESI source is mounted on an Opto-Sigma GOH-40B35 dual axis goniometer. Photos of the setup can be found in the Supplemental Figs. \ref{gonipic1} and \ref{gonipic2}, and a video of the goniometer being manipulated in both dimensions can be found at Ref. \citenum{ulibarri_2024_13621591}. This dual-stage goniometer provides rotation in two angular dimensions about a critical point 35 mm above the surface. If the source had been mounted on a simple rotation stage, attempting to adjust for high-off-center emission plumes would change the location of the physical emission site. By constructing the ESI source such that the emission site is located at the critical rotation point of the goniometer, adjusting the angle of the beam does not alter the physical location of the emission point. This means that even if the plume is firing wildly off-axis, after aiming it will fire through the center of any downstream ion optics or electrostatic gates. While existing studies may achieve this by placing the emission site directly over the center of a rotation stage, the dual axis goniometer enables this type of rotation in the two dimensions simultaneously.

A PEEK plastic sheet is machined to provide exactly the standoff needed to bring the center of the extractor aperture to the goniometer's rotation point of 35 mm above its surface. The PEEK also provides electrical isolation from the goniometer itself, which is grounded. Further discussion about the choice of this particular goniometer can be found in Section \ref{discussion}.

Rotational feedthroughs installed on the vacuum chamber are connected via flexible shafts to the goniometer knobs to provide in-vacuo angle manipulation. A minor note here is that the feedthroughs selected for this have a locking ability, as in some cases where the flexible shafts are not able to take a simple and direct line to their control, they may twist and exert a non-negligible amount of counter-torque in certain orientations. The feedthroughs can be manipulated to the desired angle, and then a set screw locks their position in place to resist any counter-torque if needed.

\subsection{Current Detecting Electrode}
An optional feature of the system described here is a downstream current-detecting electrode. A photo of the electrode is shown in the Supplemental Fig. \ref{dumbplate}. It consists of two aluminum discs mounted on a linear translation feedthrough at the top flange of a CF tee at the end of the flight tube. The linear feedthrough allows the assembly to be rapidly lowered into place and removed when desired. The smaller disc's diameter is slightly smaller than the MCP's detection area, while the larger disc's diameter is only slightly smaller than the flight tube inner diameter. 

The smaller disc can be used as a diagnostic tool to assess MCP health, as it can be used to quantify the amount of current actually reaching it. External to the chamber, the two discs can be electrically connected to provide a quick way to gather the maximum possible current yield  (to improve SNR) reaching the end of the flight tube. This method allows for rapid aiming of the goniometric system to find the optimal angle for putting the most current on target at the end of the flight tube. The larger disc's edge is covered in Kapton tape so that it does not become grounded if it accidentally is lowered too much into the flight tube. 

The electrode plate can be very useful when used as a diagnostic for the MCP, and it may also be useful for diagnostic devices or other targets where it is difficult to determine if the beam is accurately centered. However, in practice with the setup described in this study, the MCP itself can be used instead of the electrode plate to determine when the current reaching it is maximized. Nonetheless, the electrode plate can be useful as a shield to protect the fragile MCP when it is not in use. This prevents or reduces stray droplets or other ejecta from other experiments from impinging on and potentially damaging the MCP.

\subsection{Experimental Method}
After loading and installing, $\phi_{acc}$ was slowly ramped up to potential, with stable firing observed at 1800 V at a current output of around 300 nA. At lower potentials, lower current was observed, but the source behaved erratically, firing in clear discrete bursts. The source was fired at 1800 V for several minutes to ensure that output was relatively stable, with the electrode plate blocking the MCP to protect it. The electrode plate was then pulled out of the way, and the MCP was turned on to 1 kV. It was confirmed that signal clipping was occurring due to the current output from the MCP being outside the bounds of the TDC-30 input range. The MCP voltage was then lowered to 800 V to recover the full signal, and the goniometer was manipulated to maximize the voltage detected on the oscilloscope. During this two-dimensional scan, no evidence of a second emission site was observed. The pitch was then held constant at 12.5$^{\circ}$, where a signal maximum was found, while the yaw was scanned by hand across the goniometer's largest effective angular range of 22$^{\circ}$. 

At each stop, signals observed on the oscilloscope were saved as a single averaged TOF curve, the current was read off of the SMU by eye and recorded, and the yaw angle recorded by a photograph of the goniometer's scale.

Over the course of the experiment, the output current of the source dropped from 312 nA to 257 nA as the yaw was swept from negative to positive yaw. The currents reported here are thus scaled by the measured output current at each yaw angle to eliminate any effect from this current drop. As mentioned in Section \ref{discussion}, this scaling does not have a significant effect on the results.

\subsection{Data Processing}

\begin{figure}
	\centering
		\includegraphics[width=1\linewidth]{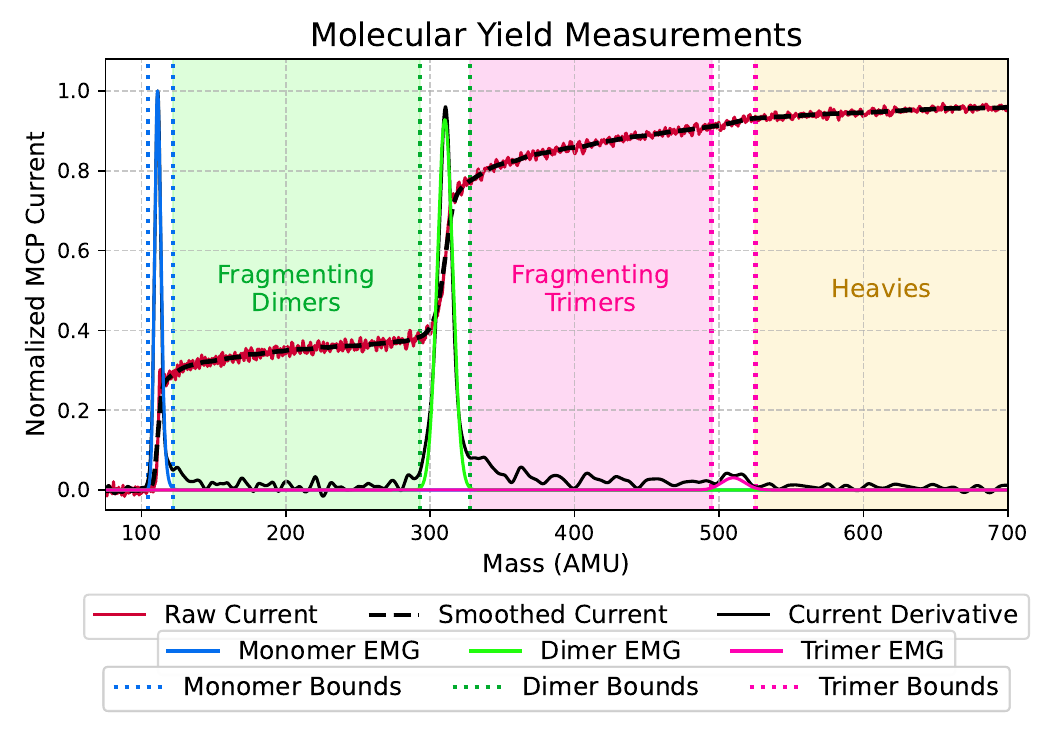}
	\caption{An example plot of a TOF curve of EMI-BF$_4$ with a plume center yaw of -11.5 $^\circ$ detailing the method of current yield measurements. The red line shows the normalized current from the MCP, the black dashed line shows the current after Gaussian blurring, and the solid black line shows the smoothed current derivative. Blue, green, and pink solid lines show the fitted EMGs to the monomers, dimers, and trimers respectively. Dotted lines show where these EMGs cross an arbitrary threshold, taken to be the bounds for those species. The change in current between these bounds was taken as the ion yield for each species. The shaded green and pink regions denote the bounds between the primary species, where fragments are created in the acceleration column and thus have m/z values that do not correspond directly to their parent species. The current rise across these regions is taken as the fragment ion yield. Lastly, the shaded yellow region denotes the heavy ion region, where droplets or other species arrive at the MCP. The current rise from the beginning of this region to the end of the mass spectrum near 2800 AMU is taken as the heavy ion yield.}
	\label{fits}
\end{figure}

After collection, the TOF curves were smoothed with a Gaussian blur to enable fitting of Exponentially modified Gaussians (EMGs) to the curve derivatives. Most of the spectra were smoothed with a blur of width 25, while the three highest angle spectra in both positive and negative yaw (representing angles greater than $\pm 14.5^\circ$ off of the plume center) were of such low SNR that they were blurred variously by widths of 50-100. The blur factor of 25 was chosen because it enabled repeatable fits across the entire dataset, excepting these highest off-center shots. The shots that require additional blurring are not used for relative ion yields due to this increased blurring. 

Fig. \ref{fits} details how the composition of each TOF curve was determined. The red solid line shows the raw current signal, while the black dashed line shows the signal after Gaussian blurring. At first glance, it may be difficult to see this dashed line because of how well it tracks the raw data. The solid black line shows the derivative of the smoothed current data. EMGs of the form 
\begin{equation}
    a_0 \frac{\lambda}{2}e^{\frac{\lambda}{2} \left(2 \mu + \lambda \sigma^2 - 2t \right)} \cdot \erfc\left(\frac{\mu +\lambda \sigma^2 -t}{\sqrt{2}\sigma}\right)
\end{equation} 
were fitted to the derivative mass lines in each spectrum, similar to the approach detailed further in Ref. \citenum{ulibarri2023detection}. These EMGs use the complementary error function $\erfc$ and have amplitude $a_0$, exponential decay term $\lambda$ , variance $\sigma ^2$, and mean $\mu$. Solid lines in Fig. \ref{fits} show the EMGs fitted to the mass species. These EMGs were used to define the conversion of the x-axis from the time domain to the mass domain based on the known monomer and dimer masses.

An arbitrary threshold of 5E-2 was selected to determine where the fitted EMGs start and stop. The index where the EMGs fitted to the monomers, dimers, and, where possible, trimers crossed this threshold was taken as the bounds of those species, shown as dotted lines in Fig. \ref{fits}. The average value of these indices was calculated across the entire dataset, and these average values were then used as the bounds for each TOF curve. The current at the lower bound was subtracted from the current at the upper bound to determine the change in current across the species, and this was taken as the ion yield of that species. The rise in current between the primary species was taken as the fragmenting or secondary species. That is, the rise in current between the upper monomer bound and the lower dimer bound was taken as the ion yield of the fragmenting dimers. The total current yield was calculated by subtracting the current from an arbitrary point near mass zero AMU from the current near the end of the recorded data near mass 2800 AMU. These arbitrary points were selected because the high voltage gate pulses create small transient signals in the current data at the beginning and end of the spectra.

It is important to note that the trimer yield is reported using these bounds, but a successful fit of the trimer EMG was only found in six of the 38 spectra. Thus, the trimer yield is reported as the current rise across this boundary regardless of presence of a visually identifiable trimer feature. 

Because the scope performed the averaging of the spectra and only one average was taken at each yaw angle, there is no straightforward way to calculate a standard deviation to generate a standard error of means. However, on a previous experimental run, four repeated measurements were taken at a single yaw value roughly corresponding to 10$^\circ$ off-plume center. The data from this run was later abandoned because the center of the plume was not correctly measured due the high currents generated by the goniometric system and the limitations of the amplifier and MCP setup, but these four data points were not affected by those limitations since the current was much lower at the roughly 10$^\circ$ off-center. This repeated measurement yielded a standard error of means of 1-2$\%$ of the ion yields, except for the trimer measurement, which had an error of 10$\%$. We conservatively apply the maximum error of 2$\%$ taken at the maximum dimer yield as a proxy error across the entire dataset. This likely overestimates error in the center of the plume while underestimating it at the fringes, but in the absence of true error measurement it is a useful proxy error estimate.

\section{Results}
\begin{figure}
	\centering
		\includegraphics[width=1\linewidth]{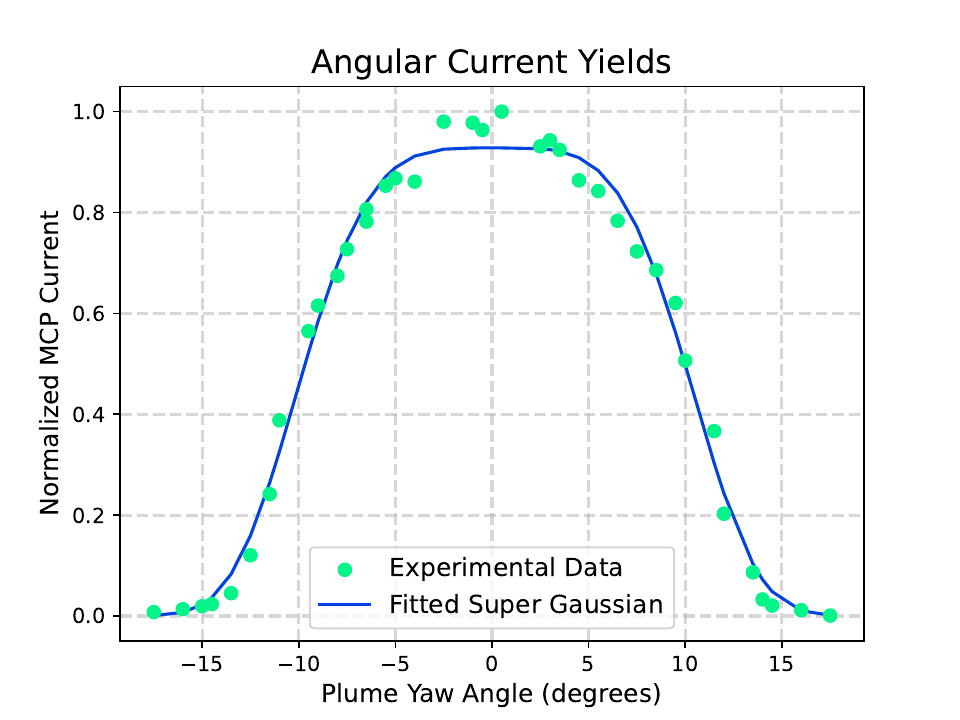}
	\caption{Normalized current yields on the 42 mm Ø microchannel plate detector 1 meter away from the emitter. The x-axis shows yaw angle with respect to the beam center line, which was found to be 4.7$^{\circ}$ off the emitter geometry in yaw. This entire scan was taken at a pitch angle of 12.5$^{\circ}$, which was empirically found to be the beam center line's pitch. A super-Gaussian fit of the data yields characteristic width $\theta_0$ of 11$^{\circ}$, a sharpness $n$ of 2.1, and an 80$\%$ divergence half angle of 8.7$^{\circ}$.}
	\label{supergaussfig}
\end{figure}

\begin{figure*}
	\centering
		\includegraphics[width=1\textwidth]{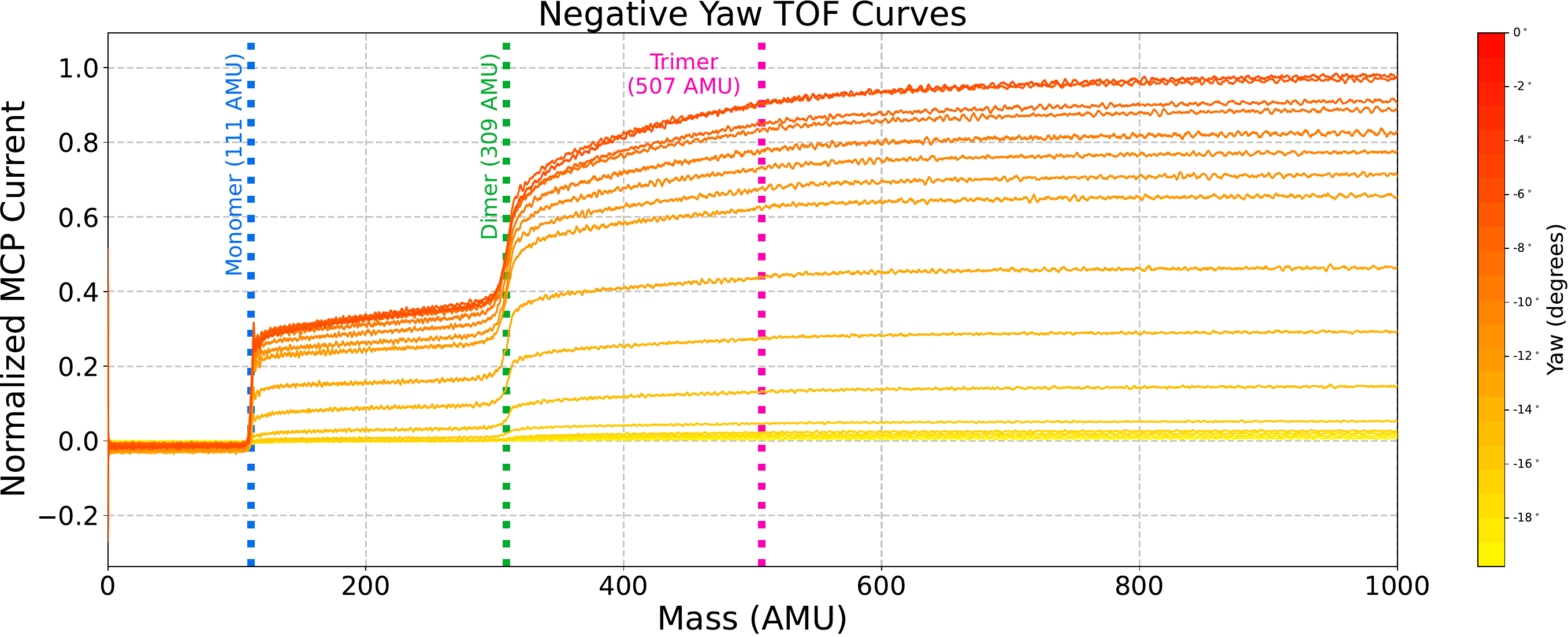}
	\caption{TOF curves from the negative yaw (with respect to the plume center line, not the emitter geometry) scans. The positive yaw scans are excluded for readability. The major steps are found at the monomer, EMI$^+$, at mass 111 AMU and the dimer, (EMI-BF$_4$)EMI$^+$, at mass 309 AMU. The trimer feature is only visible by eye in a handful of spectra, but can be more easily seen in Fig. \ref{supergaussfig}. These plots have been normalized to the highest data point within this plot window, which is limited to 1000 AMU to more easily see structure. However, the data extends to about 2800 AMU, where measurements are made to calculate the total current yield and the heavy ion yield. Note also that these curves have been baseline corrected to subtract constant background current, which was also found to be angle-dependent.}
	\label{multishot_neg}

\bigskip

		\includegraphics[width=1\textwidth]{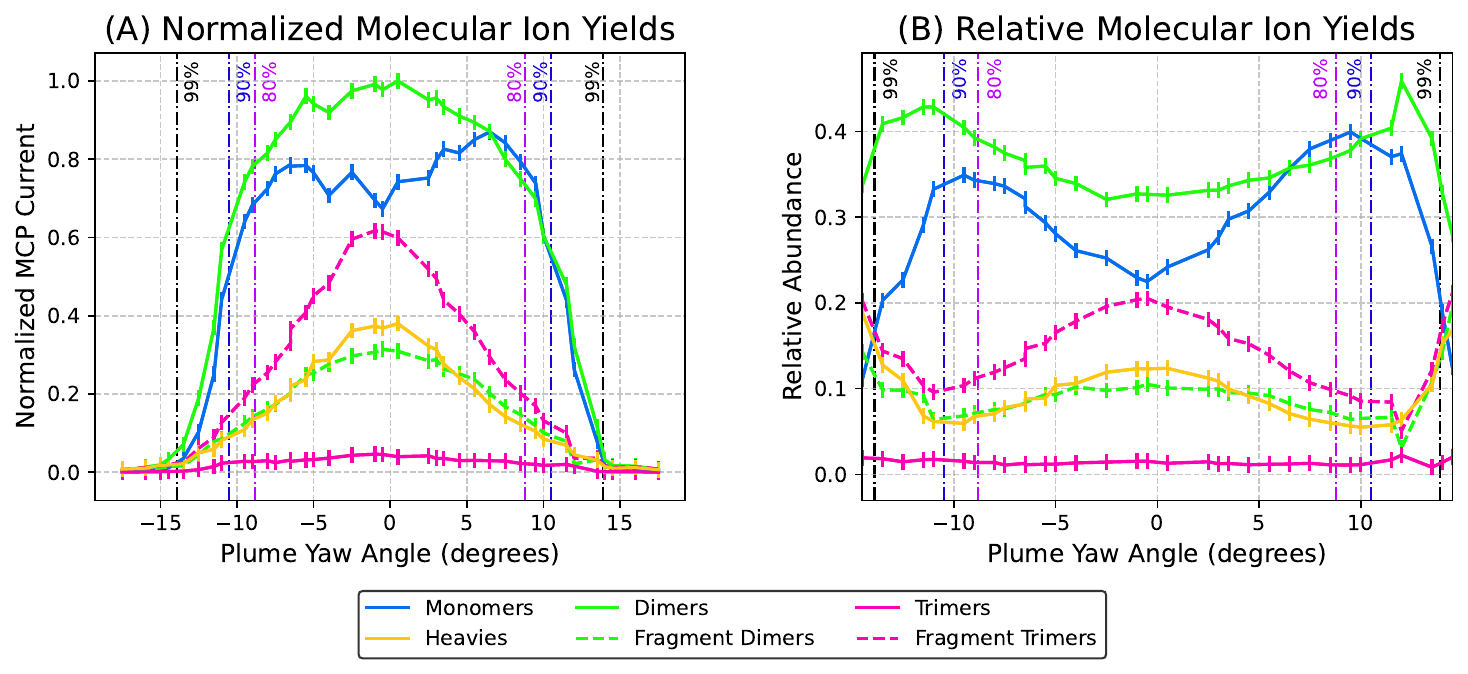}
	\caption{(A) Molecular yields of the the monomers (blue), dimers (green), trimers (pink), fragments thereof (dashed lines), and high mass species (yellow) normalized to the maximum, found to be the dimer yield near the center of the plume. The super-Gaussian fit in Fig \ref{supergaussfig} was used to determine divergence angles that contain 80$\%$, 90$\%$, and 99$\%$ of the current. These are shown in vertical dash-dot lines. (B) Relative abundances of the monomers (blue), dimers (green), trimers (pink), fragments thereof (dashed lines), and high mass species (yellow). Note that the x-axis range is slightly smaller than that of (A) to eliminate noisy data where the SNR drops considerably. The monomers and dimers show relative minima near the center, where the trimers and heavies show maxima there. However, these trends reverse near the edges, creating a characteristic `bow tie' shape.}
	\label{absrel}
\end{figure*} 
%\clearpage  % Force a page break to ensure figures stay on the same page
Total current yields are shown in Fig. \ref{supergaussfig}. Following Ref. \citenum{petro2022multiscale}, a super-Gaussian of the form 
\begin{equation}
    A exp\left[-\left(\frac{(\theta-\theta_c)^2}{\theta_0^2}\right)^n\right],
\end{equation} was fitted to the data, where $\theta$ is the independent angle, $\theta_c$ is the angle of the plume's center line, $\theta_0$ is a beam width parameter, $n$ is a sharpness parameter, and $A$ is the amplitude. The sharpness parameter $n$ was found to be 2.2 and $\theta_0$ was found to be 11.0$^{\circ}$. $\theta_c$ was found to be off the emitter's center line by 4.7$^\circ$. This 4.7$^\circ$ yaw angle is taken to be the emission plume center yaw angle, and all subsequent yaw measurements are reported as deviations from this plume center, rounded to the nearest half-integer as this was the original resolution recorded off of the goniometer scale. As in Ref. \citenum{petro2022multiscale}, divergence angles are calculated from the super-Gaussian fit as the half angles which contain the specified percentage of current. The super-Gaussian yields an 80$\%$ divergence half angle of 8.7$^\circ$, a 90$\%$ half angle of 10.4$^\circ$, and a 99$\%$ half angle of 13.8$^\circ$.

Fig. \ref{multishot_neg} shows the TOF curves taken at various negative yaw angles from the plume center line at 0$^\circ$ (4.7$^\circ$ with respect to the emitter center line) up to the greatest negative yaw -17.5$^\circ$. The positive yaw lines have been excluded for clarity, but a similar plot with the positive lines is included in the Supplemental Fig. \ref{multishot_pos}. It can easily be seen that the overall collected current drops off at higher plume angles as expected. To the bare eye, the rough shapes of the blurred TOF curves do not exhibit wildly different structures, although as we will see, they produce markedly different molecular yields. However, trimer features can be faintly identified by eye in a handful of spectra predominantly near about 10$^\circ$.

Fig. \ref{absrel} (A) shows the individual molecular yields as a fraction of the dominant species, the dimer yield just left of center of the plume, as a function of yaw angle. Fig. \ref{absrel} (B) shows the relative ion yield of each individual species, defined as the ratio of the ion yield of each species divided by the total sum of ions. Because of the low current observed at high angles, this plot exhibits erratic fluctuations beyond $\pm 14.5^\circ$, which were also the regions where the Gaussian blur had to be increased to recover EMG fits. Thus, the regions beyond $\pm 14.5^\circ$ have been excluded for clarity.

Monomers and dimers are found to be at a relative minimum at the center of the plume, while fragmenting trimers and heavies are found to be at maximum there. The yields are largely symmetric, but while the dimers are consistently dominant in the negative yaw direction, the monomers are slightly dominant between about $+6.5^\circ$ and $+10^\circ$. While the plots in (A) show slightly erratic behavior in the center region, with the monomer line in particular showing fluctuation, the relative data in (B) shows clear trends that are largely symmetric across the yaw range of $\pm 8^\circ$, excepting the aforementioned monomer dominance at high positive yaws. The relative yields creates a `bow-tie' shape between the monomers and the trimer fragments as the monomers increase towards the edges and trimers decrease, but this reverses near the plume edges.

A CSV file containing the yields for each yaw angle is provided in the Supplemental.

\section{Discussion}
\label{discussion}

\subsection{Population Yields}
\emph{Jia-Richards 2024} \cite{jia2024numerical} chose to combine fragmenting species with their parent molecules for the purposes of comparison. The Supplemental Fig. \ref{absrelcomb} shows a version of Fig. \ref{absrel} (A) and (B) using such a method, where the fragmenting dimers are included with the dimer population and the fragmenting trimers are included with the trimer population. The trends are similar in Figs. \ref{absrelcomb} and \ref{absrel}, except that the combined dimers remain dominant across the entire plot range rather than dipping below the monomers near yaw +6$^\circ$. Whereas \emph{Jia-Richards} found dimers to peak in the center of the plume, we find dimers to be at a relative minimum at the center of the plume, although when they are combined with their fragments as in Fig. \ref{absrelcomb}, this line is largely flat until the extremities of the plume beyond $\pm 12^\circ$.

\emph{Jia-Richards 2024} \cite{jia2024numerical} also found trimers and heavies to be roughly equal across the plume, which terminates at $\pm 6^\circ$, and both to be at maximum at the center. We similarly find heavies and trimers to be at maximum in the center, but that trimers, or more specifically fragmenting trimers, are significantly more common than heavies across the scan range. However, the plume in our study goes well beyond $6^\circ$, and we find that at around $11^\circ$, the relative abundance of monomers and dimers begins to drop while that of heavies and trimers begins to rise, thus creating a `bow tie' shape between the monomers and trimers/heavies that is not observed in \emph{Jia-Richards 2024} \cite{jia2024numerical}.

\emph{Schroeder et. al 2023} \cite{schroeder2023angular}, from which \emph{Jia-Richard 2024} drew data, performed angular studies focused on the voltage dependence of the ion beam and found much narrower beam profiles than are observed here. Specifically, \emph{Schroeder et. al 2023} found a measured spread (the full angle over which the beam can be distinguished from background) to be 9-11$^\circ$ over the voltage range of 800-950 volts, as compared to the 32$^\circ$ found in this study (although if we use the more conservative definition of beam width defined by the 99$\%$ divergence angle, our plume width would be 28$^\circ$). Whereas \emph{Schroeder et. al}'s emitter fired 10 nA at 800 V and 40 nA at 950 V, the work presented here fired 257-312 nA over the course of the experiment, all at 1800 V. As \emph{Schroeder et. al} note, the beam divergence angle increases with output current, as the increasing field strength in the center of the plume forces ions out in a radial direction.

\begin{figure}
	\centering
		\includegraphics[width=1\linewidth]{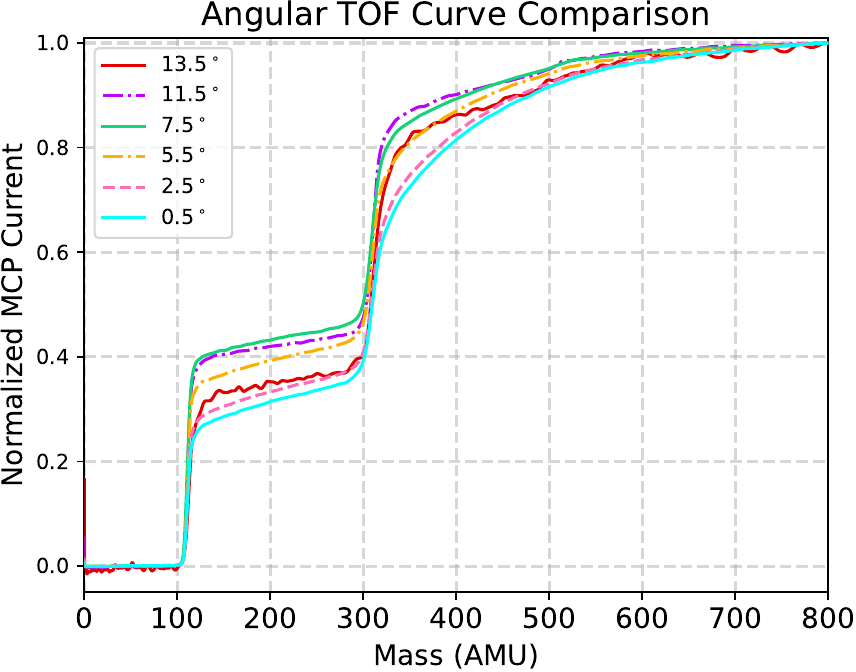}
	\caption{A comparison of the positive yaw TOF curves normalized to the maximum current out to 800 AMU. Higher lines indicate a greater proportion of the current being derived from low-mass molecules, and thus higher lines mean lower average mass of the plume. As the plume angle increases, the lines shift up, indicating lower average mass of the molecular constituents. Between 11.5 and 13.5$^\circ$, the average mass begins trending up again.}
	\label{comparison}
\end{figure}

\emph{Schroeder et. al 2023} also found that at low accelerating potential (800 V), the center of the beam had higher average molecular mass than the edges, whereas at high potential (950 V), the inverse was true and the center of the beam featured smaller average mass than the edges. While the work presented here only considers a single acceleration potential of 1800 V, we find that with our plume the highest average mass is near the center, and it decreases with plume angle until about 12$^\circ$, after which it begins to rise again. This can be seen in Fig. \ref{comparison} by comparing the TOF curves normalized to the plot window. Higher lines indicate a greater proportion of the current being derived from low-mass molecules, and thus higher lines mean lower average mass of the plume. It can be seen that the scan at 0.5$^\circ$ is the lowest of the group, and the line rises with each angle until 11.5$^\circ$. At 13.5$^\circ$, however, the line has dropped back down nearly to the level of the plume center. While Fig. \ref{absrel} clearly shows increased relative fragment abundance at the center of the plume, this can be qualitatively observed in Fig. \ref{comparison} by noting that the two smallest angles, 0.5$^\circ$ and 2.5$^\circ$, both have much more rounded shoulders between 300 and 400 AMU; that is, instead of the comparatively sharp change in current near 330 AMU shown in the other data, these two center data points exhibit much less well-defined boundaries between the dimer yield and the fragmenting trimer yield. Fig. \ref{comparison} shows positive yaw data, but similar trends occur in the negative yaw.

Compared to the externally wetted tungsten needle with an extractor aperture of approximately 3000 $\mu$m Ø used in this study, \emph{Schroeder et. al} used a porous emitter tip with a 300 $\mu$m Ø extractor aperture. These discrepancies, combined with the order of magnitude difference in output current, all provide ample reasoning for the differences observed between the studies. Nonetheless, \emph{Schroeder et. al 2023} shows the importance of voltage in beam composition, and future studies with our system may be performed to better understand these dynamics.

In the work presented here, the output current of the emitter fell from 312 nA at the start of the scan in negative yaw and fell to 257 nA at the end of the scan at positive yaw. To remove this from the reported trends, the current measurements in Fig. \ref{supergaussfig} were scaled by the output current before normalization. This changes the super-Gaussian fit slightly, with the center angle rising from 4.5$^\circ$ to 4.7$^\circ$, but does not change any other super-Gaussian parameter to the reported number of significant digits. The scaling similarly has a small effect in Fig. \ref{absrel} (A), shifting the left shoulder of all species down slightly, but it does not affect Fig. \ref{absrel} (B), which only shows relative abundances of the molecular species. However, while the scaling itself has no effect on the relative yields \emph{calculation}, it is quite likely that the reduced current has an effect on the molecular composition of the plume. As discussed in \emph{Petro et al. 2022} \cite{petro2022multiscale}, higher currents yield higher plume densities and thus impact-associated fragmentations (including higher monomer yield if the fragmentation occurs early enough in the acceleration column). However, the results that we see here are the opposite, as an increased monomer yield is found in the positive yaw, where current was minimal. Even so, the change in current was small and slow, and so it is not surprising that a measurable effect was not observed.

\subsection{The Importance of Two-Dimensional Steering for Angular Plume Studies}
\label{2dmol}

We can be reasonably sure that the yaw scan presented here did in fact pass through the conical beam center based on the signal levels detected during initial zeroing. Note, however, that the beam center line was 12.5$^\circ$ in pitch (which it might be recalled is the axis that is traditionally unavailable to systems that use a rotation stage instead of a dual-stage goniometer). Consider how different the plume is at 12.5$^\circ$ (or any other arbitrary non-zero angle) in Figs. \ref{supergaussfig} and \ref{absrel} than it is at the center. Even scanning across to the maximum yield in yaw, only about 18$\%$ of the current maximum would have been received by the MCP. Worse, the molecular yield would have wildly underestimated the fragment, trimer, and heavy populations while overestimating the monomer and dimer populations.

These results also show the danger of studying ESI plumes only in a single dimension, as a failure to account for the second dimension can wildly skew experimental results and lead to data that is simply incorrect or misrepresentative. It should again be considered that a key finding of \emph{Schroeder et. al 2023}\cite{schroeder2023angular} was that the plume center wandered in the one measured dimension from +7$^\circ$ to -8$^\circ$ as a function of acceleration over the range of 800-950 V. If such a 15$^\circ$ variation is possible in one dimension, in the same experiment no less, across just a 15$\%$ variation in voltage, then clearly there is a great danger in only varying one of the two dimensions of the conical beam angle, and there is yet more danger in failing to have any angular variation at all and simply assuming that the plume is firing on-axis to the emitter geometry.

Even so, in this work we have only scanned across the plume center line in the yaw direction while the pitch was held constant.  However, it should not necessarily be assumed that the results presented here are axisymmetric and can be taken to be equally true for the pitch angle. It may very well be that small imperfections in the needle or the extractor may result in differences between the pitch and yaw, which is to say that the plume width and molecular composition in the pitch axis may not necessarily mirror that of the yaw axis. Future studies will target two-dimensional scans to create a complete picture of the plume structure. However, such an experiment would be time-consuming and difficult to conduct by hand, as this study was. A computer-controlled motorized setup is thus being constructed to enable software-defined scans and automated data taking, which will rapidly speed the experiment up and reduce human error.

\subsection{Two-Dimensional Steering as Lab Utility}

While this work has focused on the application of direct aiming for angular TOF studies, it should also be noted that the ability to manually walk ions onto various targets is incredibly useful in everyday laboratory experiments. The dramatic increases in SNR enabled by direct aiming have made the goniometer-based system an integral part of our laboratory, and it is now a critical aspect of virtually all of our ongoing projects \cite{bell2024electrospray,d2024characterization,hofheins2024electrospray,geiger2022energy}.

Indeed, as mentioned above the beam center line angle of 4.7$^{\circ}$ in yaw and 12.5$^{\circ}$ in pitch would have yielded a maximum of 18$\%$ of the current on the MCP if only yaw rotation were available, and even less had the system been incapable of rotation at all. That is, had it not been for the goniometric system, any targets or diagnostics placed in the emitter axis line would receive only a small fraction of the emitted current. With ESI sources, it is not unheard of that entire experiments would be scrapped after source installation and vacuum pumpdown simply because the emission line was so wildly off-center that appreciable or high-SNR signals could not be detected downstream. Thus, the power of the goniometric system is exemplified with this experiment; without the goniometer the entire experiment would have at best produced weak signals far from the beam center line, and in all likelihood the experiment would have been scrapped. However, with the goniometer, the beam was walked onto the downstream targets, and not only was the experiment thus salvaged from the off-angle firing, but maximal SNR scientific data was acquired in the process. 

The dramatic increase in SNR enabled by the 2-D goniometric aiming even created unexpected problems because the observed currents on downstream detectors were not designed to handle such high throughput. The TDC-30 transimpedance amplifier is limited to a 4 $\mu$A input, which produces a signal rail at 2 V. This limit was consistently reached by our system when operating near the true beam center line, and so as described in Section \ref{TOF}, the MCP voltage was given an amplification voltage of 800 V rather than the nominal 1 kV. In a way this illustrates the power of the 2-D goniometric system and its utility in the laboratory. Our electronics, designed to resolve minute signals through high amplification, proved to be too sensitive for the increased SNR afforded by the goniometric capability.

The unfortunate effect of this reduced MCP amplification voltage is that the absolute current readings are unreliable. Hamamatsu was unable to give us gain characteristics for the MCP at reduced voltage, and so we are only able to report the current levels in Figures \ref{absrel} and \ref{supergaussfig} as normalized measurements rather than absolute current measurements.

\subsection{On the Choice of Goniometer}
The GOH-40B35 goniometer employed in this experiment is not designed for ultra-high vacuum use, as it uses brass and black oxide metallic surfaces. Some companies offer goniometers with stainless steel design that is UHV-compatible, but these tend to be much more expensive, much more limited in terms of design variations (typically offering less angular range), and have much longer lead times. We chose the GOH-40B35 goniometer because its critical rotation point was high, at 35 mm, giving flexibility to the design of the ESI sources, and because it has a large maximum effective throw distance of $\pm22^{\circ}$ in one dimension (yaw in our experiment) and $\pm17^{\circ}$ in the other (pitch in our experiment). These wide angle ranges allow for maximum flexibility in the experiment. The lubricants internal to the goniometer are likely the most critical aspect of vacuum compatibility, as these may outgas considerably. However, Opto-Sigma (and a number of other companies we spoke to during selection) allow for use of vacuum-rated lubricants in the non-vacuum goniometers at an additional fee. As we perform our studies in the low E$^{-5}$ or high E$^{-6}$ Torr range, the suboptimal UHV characteristics of the goniometer loaded with vacuum grease have not been an issue.

The GOH-40B35 is a dual-axis goniometer with a large range in both dimensions. However, it is also large and heavy as a result. An alternative design has also been made using a rotation stage to enable, in principle, full 360$^\circ$ rotation in yaw, but a single-stage goniometer to enable pitch control \cite{geiger2024secondary}. This enables use of much lighter and smaller ESI setups, and enables greater rotation in yaw, which can be used to swing the beam onto multiple instruments in a circular arc around the source without having to open the vacuum chamber to reset the experiment. However, the rotation stage itself is much larger and heavier than the goniometer, and it takes up a larger section of the vacuum chamber.

\subsection{Future Experiments}
As mentioned in Section \ref{2dmol}, we are currently assembling a motorized control system to enable software-defined movement and automated data taking. Controlling the goniometers by hand is time consuming and can lead to human error. In particular, a simple problem run into in experiments was that the human operator would accidentally rotate the wrong dimension midway through a scan, suddenly changing the pitch when it was meant to be held constant, and so the scan would have to be scrapped and started anew. Furthermore, by automating not just the movement but also the data acquisition, much less error can be induced, and in the timespace that it takes a human to take a single data point, a computerized setup may take several. This would enable true statistical analysis of the data for proper error measurement rather than the proxy error that was used in this study while also minimizing the current drift observed in this study.

In addition to the motorized two-dimensional plume TOF characterization mentioned in section \ref{2dmol}, we are currently building a setup to perform a statistical study of plume divergence angle across multiple needles and voltages. We also intend to perform studies of the ESI plumes in both negative and positive ion mode to determine if there are meaningful differences between the resultant plumes.

The data presented here was taken over a 17 minute span. A key assumption in this work is that the plume firing angle did not wander during this time. With the motorized two-dimensional setup, we will also be able to determine if and how the beam firing angle wanders over time, even as all firing parameters (e.g., $\phi_{acc}$) are held constant.

\section{Acknowledgments}
The authors would like to thank Samantha Adamski for her assistance in logging the data taken in this experiment. This work was supported in part by the Heising Simons Foundation 51 Pegasi B Fellowship grant number 2024-5175. 
This work made use of the Cornell Center for Materials Research Shared Facilities which are supported through the NSF MRSEC program (DMR-1719875).
A portion of the work was supported by a grant from the Jet Propulsion Laboratory, California Institute of Technology, under contract with the National Aeronautics and Space Administration (80NM0018D0004).

\section{References}
\nocite{*}
\bibliography{bibliography}% Produces the bibliography via BibTeX.
\appendix

\section{Appendixes}

\begin{figure}
	\centering
		\includegraphics[width=1\columnwidth]{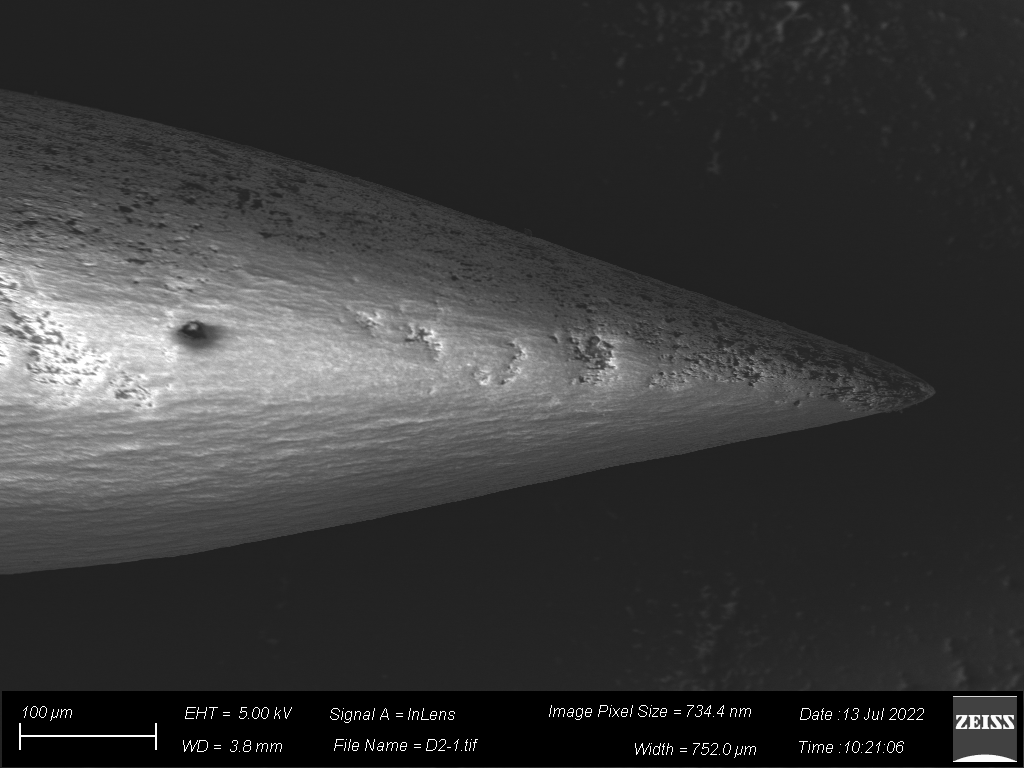}
	\caption{A wide angle scanning electron microscope image of the tungsten needle used in these studies. Taken at the Cornell Center for Materials Research Facilities. It should be noted that these images were taken at needle fabrication, and the needle had been in intermittent use for over a year before these experiments were conducted.}
	\label{needle1}
\end{figure}

\begin{figure}
	\centering
		\includegraphics[width=1\columnwidth]{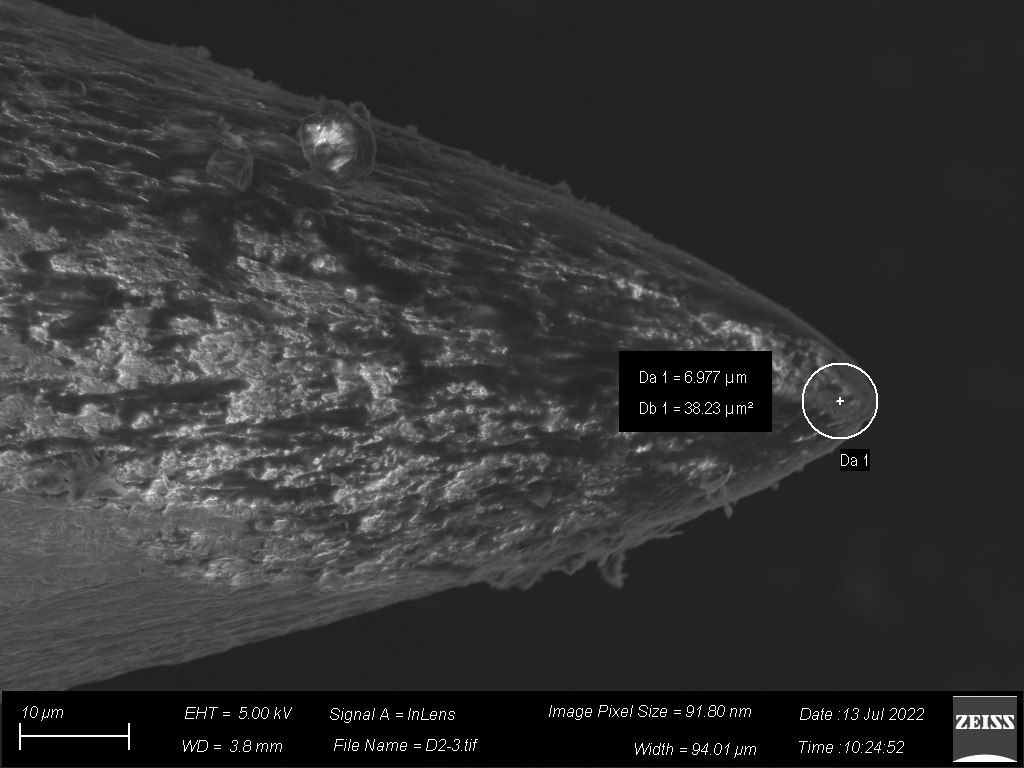}
	\caption{A scanning electron microscope image of the tungsten needle tip with a diameter measurement. Taken at the Cornell Center for Materials Research Facilities. It should be noted that these images were taken at needle fabrication, and the needle had been in intermittent use for over a year before these experiments were conducted.}
	\label{needle2}
\end{figure}

\begin{figure}
	\centering
		\includegraphics[width=1\linewidth]{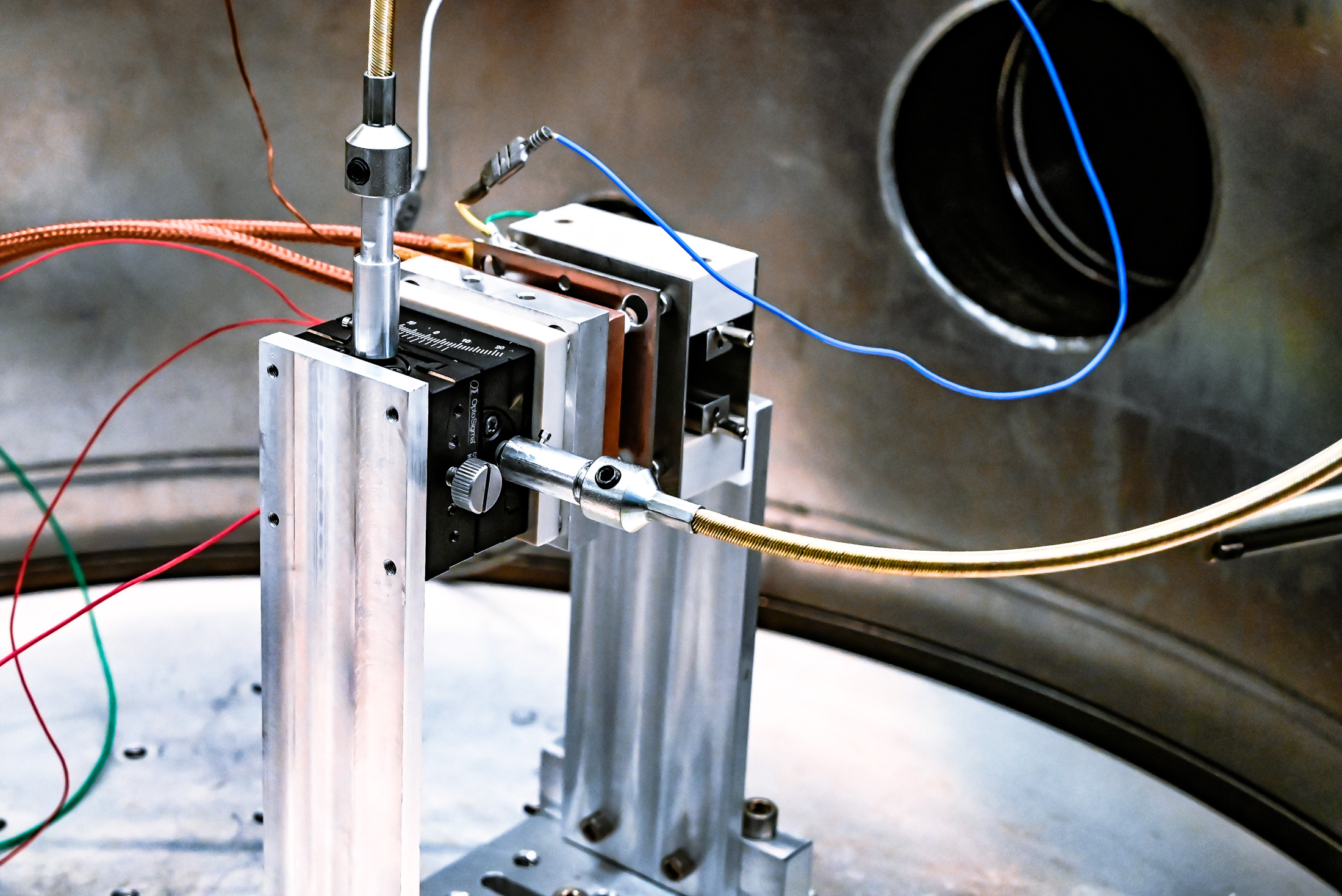}
	\caption{Image of the goniometer installed in the vacuum chamber. Flexible shafts connected to rotational feedthroughs allow in-vacuo control of the two angular dimensions. The ESI source faces the electrostatic TOF gate, and a hole leading to the TOF flight tube can be seen on the chamber wall.}
	\label{gonipic1}
\end{figure}

\begin{figure}
	\centering
		\includegraphics[width=1\linewidth]{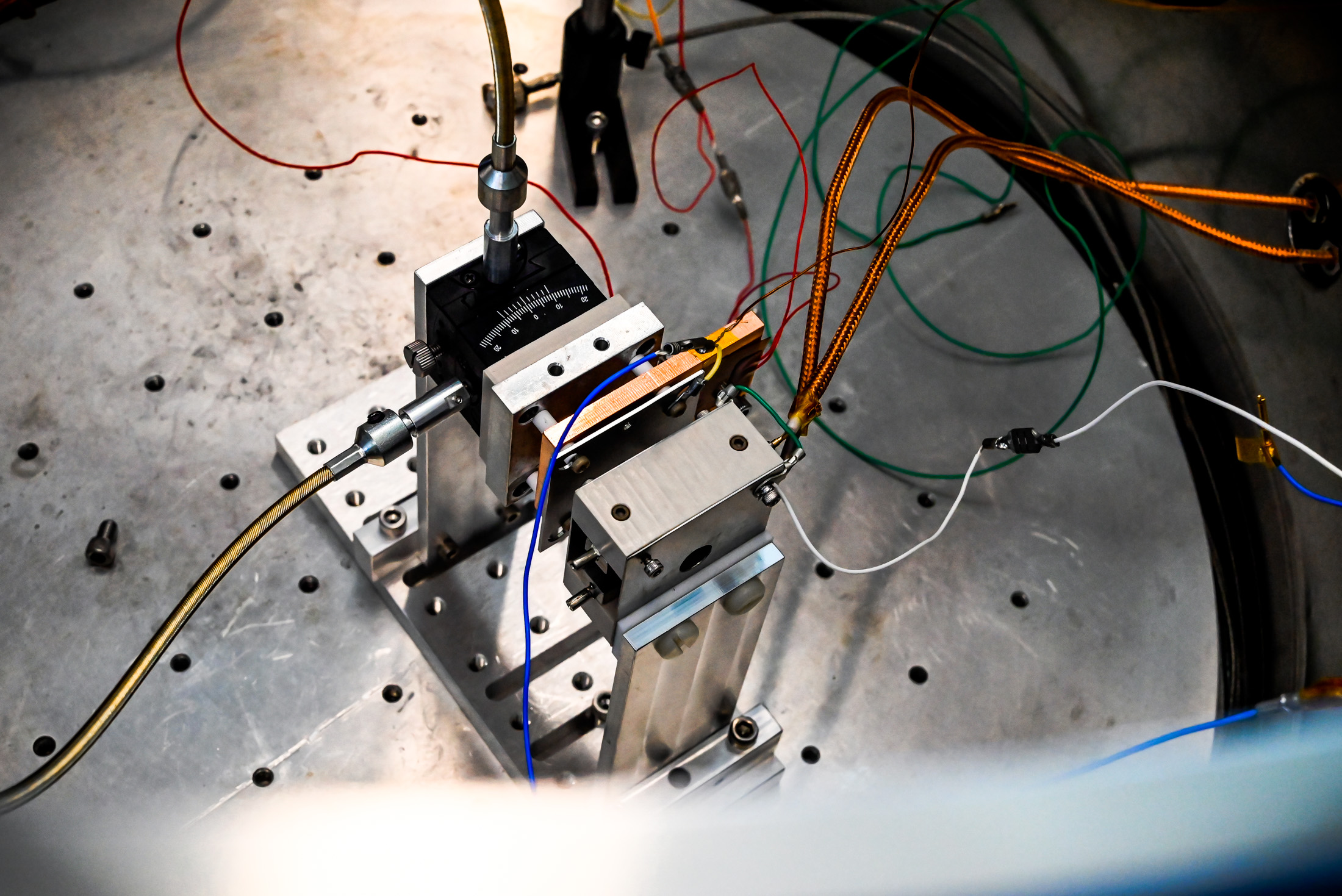}
	\caption{Photo of the 2-D goniometric system installed in the vacuum chamber. The flexible shafts can be seen connected to the black goniometer, and they allow for in-vacuo control of both pitch and yaw. In this photo, the liquid reservoir is made of copper to facilitate heating of the ionic liquid, but the heater was not activated in this study.}
	\label{gonipic2}
\end{figure}

\begin{figure}
	\centering
		\includegraphics[width=1\linewidth]{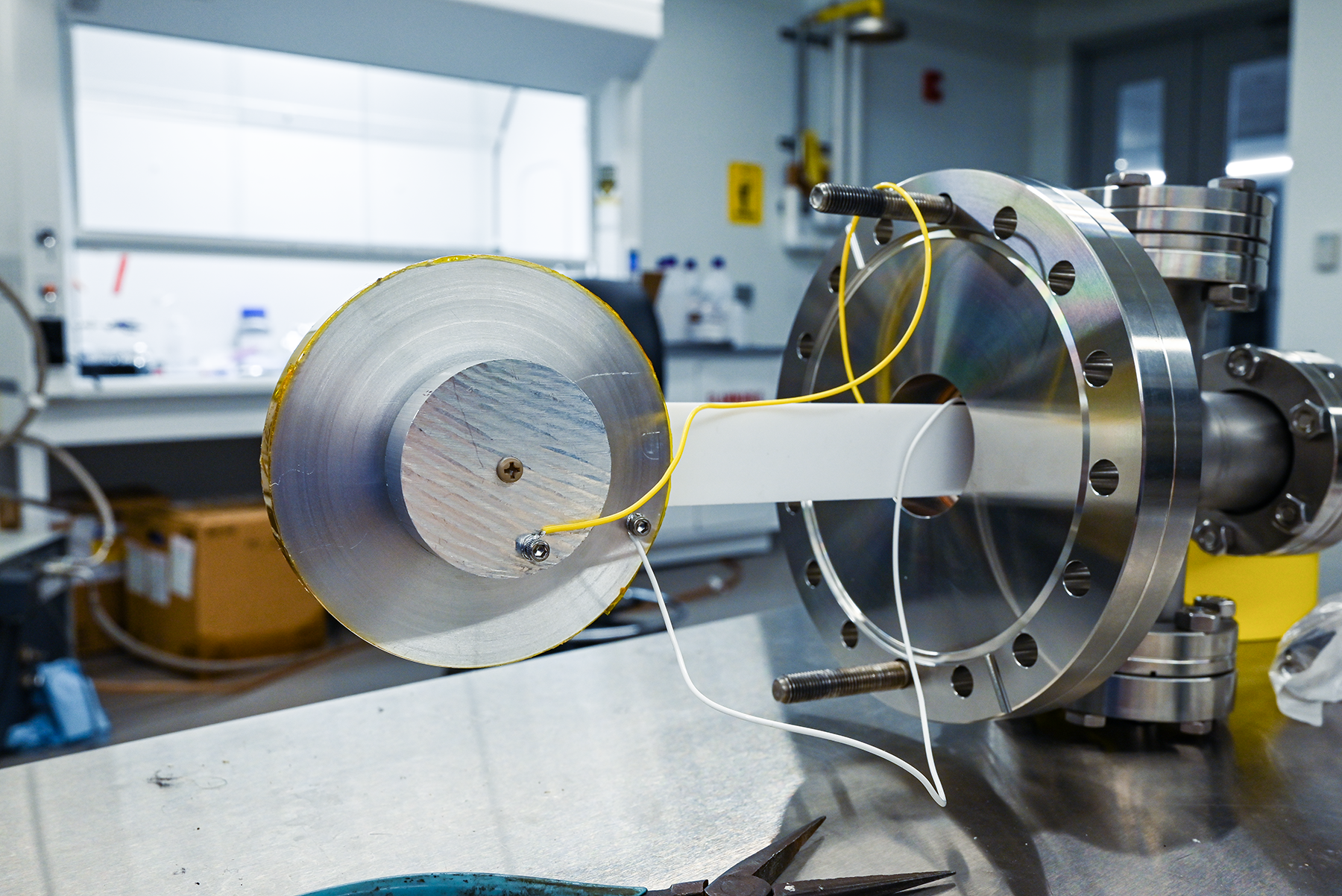}
	\caption{The downstream current detecting electrode. This simple device consists of two aluminum discs mounted on a linear translation feedthrough, allowing it to be rapidly dropped into or removed from the beamline. The larger of the two discs' outer rim is covered in Kapton tape to prevent it from being ground if it is accidentally extended too far and comes into contact with the bottom of the flight tube.}
	\label{dumbplate}
\end{figure}

\begin{figure*}
	\centering
		\includegraphics[width=1\textwidth]{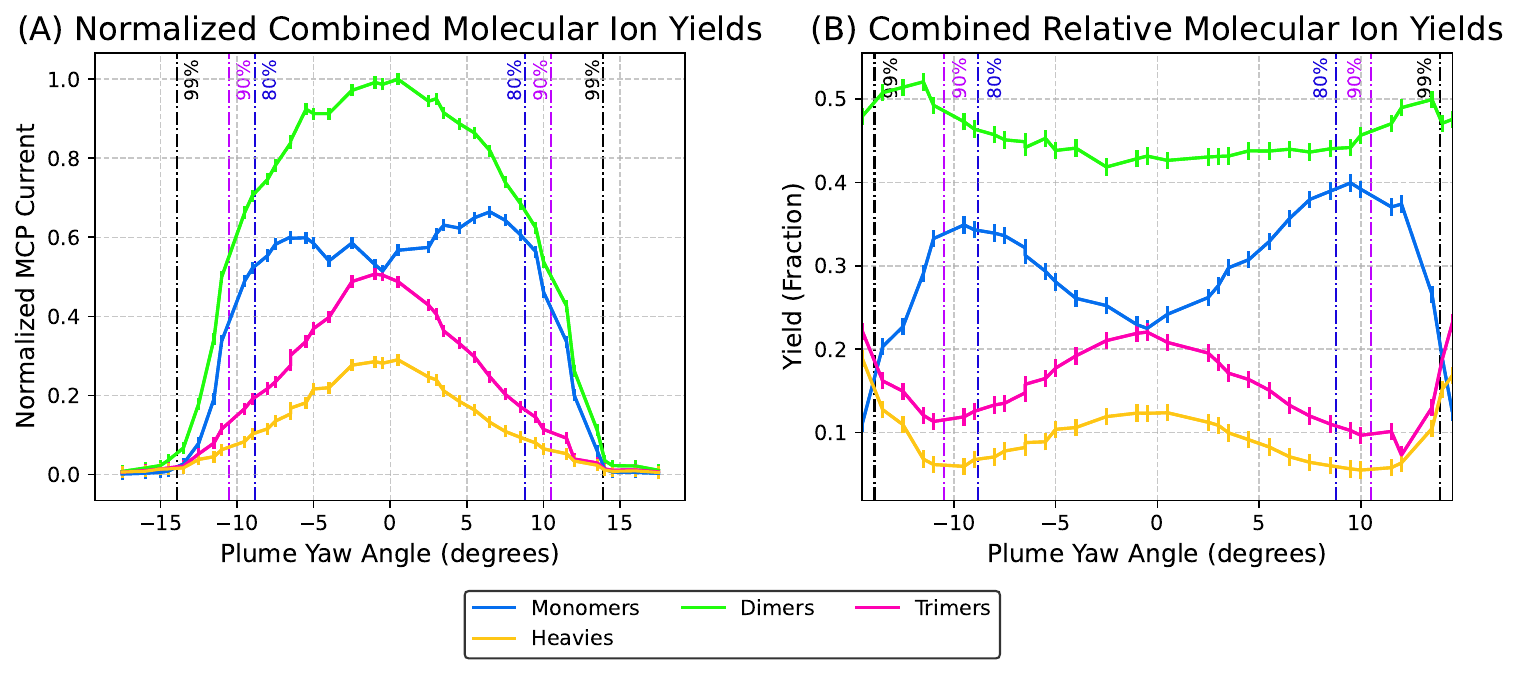}
	\caption{(A) Total molecular current yields on the 42 mm Ø microchannel plate detector 1 meter away from the emitter as in Fig. \ref{absrel}, but with fragmenting species combined with the parent molecules (that is, fragmenting dimers being included with dimers, etc.). The x-axis shows yaw angle with respect to the beam center line, not the emitter center axis, which was found to be 4.7$^{\circ}$ off in yaw. This entire scan was taken at a pitch angle of 12.5$^{\circ}$, which was empirically found to be the beam center line's pitch. (B) Relative yields of the monomers, dimers, trimers, and high mass species in the conical beam slice taken at a pitch of 12.5$^{\circ}$. As in (A), the fragmenting species have been included with the parent molecules.}
	\label{absrelcomb}
\end{figure*}

\begin{figure*}
	\centering
		\includegraphics[width=1\textwidth]{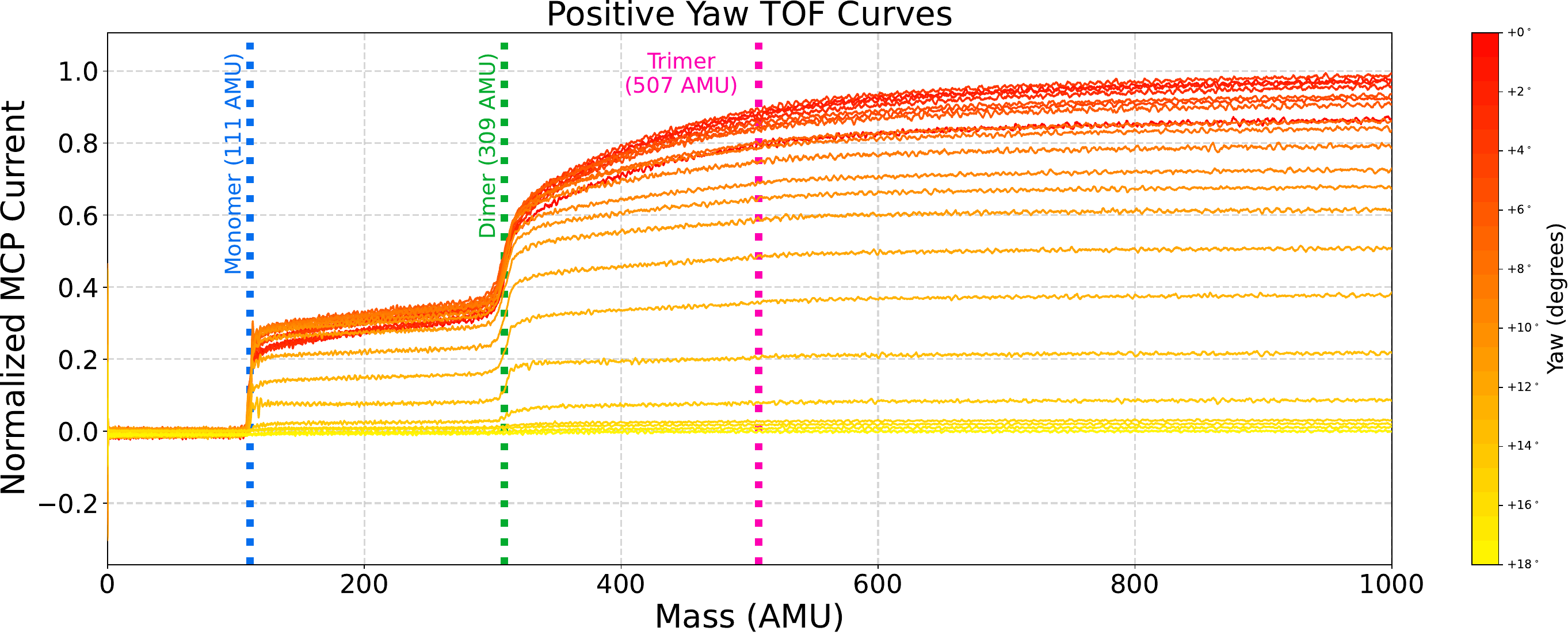}
	\caption{TOF curves from the positive yaw (with respect to the plume center line, not the emitter geometry) scans. The negative yaw scans are excluded for readability. The major steps are found at the monomer, EMI$^+$, at mass 111 amu and the dimer, (EMI-BF$_4$)EMI$^+$, at mass 309 amu. }
	\label{multishot_pos}
\end{figure*}

\end{document}